\documentclass[11pt]{article}

\usepackage[margin=1in]{geometry}
\usepackage{amsmath,amssymb,amsthm}
\usepackage{graphicx}
\usepackage{float}
\usepackage{hyperref}

\newtheorem{theorem}{Theorem}
\newtheorem{proposition}{Proposition}
\newtheorem{definition}{Definition}

\DeclareMathOperator{\Var}{Var}
\newcommand{\cx}{\preceq_{\mathrm{cx}}}
\newcommand{\R}{\mathbb{R}}
\newcommand{\E}{\mathbb{E}}

\title{Mixture-Preserving, Arbitrage-Free Interpolation\\ for Volatility-Surface Models}
\author{%
Thijs van den Berg\thanks{The author thanks Alan L. Lewis for valuable feedback and discussions.}\\
Simu Labs \quad \texttt{thijs@simu.ai}}

\date{}

\begin{document}
\maketitle

\begin{abstract}
Given risk-neutral densities of a tradeable forward, fitted as $N$-component
mixtures at a finite set of expiration pillars, we look for a continuous-time
interpolation that is (i) \emph{mixture-preserving}, remaining a mixture of the same
kernel (generically with more components than either pillar), and (ii)
\emph{arbitrage-free} across expiries. The second requirement is the \emph{peacock}
(convex-order) property, equivalently a non-negative Dupire local volatility; for
full-support kernels (Gaussian, lognormal) it gives a unique continuous
local-volatility diffusion (Lowther). We construct such an interpolation in a fixed
$2N$-component family, freezing both pillars' components and moving only their weights.
Applied to mixture term-structure models, it lifts Brigo--Mercurio to time-varying
weights and reaches the free-per-strike-width generality of SANOS at additive cost.
\end{abstract}

\section{Introduction}

Option-pricing models routinely fit the risk-neutral density of a forward at each
quoted expiry as a finite mixture of (log)normal kernels. Quoted expiries are sparse,
so pricing or hedging between them needs the density at intermediate times, and that
interpolation should do two things at once. It should be \emph{mixture-preserving},
staying inside the mixture family the densities were fitted in, so the existing
parametrization and its calibration carry over; and it should be \emph{arbitrage-free},
adding no calendar-spread arbitrage, so the interpolated surface stays
tradeable. The second condition is exactly the \emph{peacock} (convex-order) property,
and for the full-support kernels used in practice it is equivalent to a non-negative
Dupire local volatility, i.e.\ a genuine local-volatility diffusion through the data.

We give a constructive interpolation meeting both conditions. The device is a
\emph{frozen pool}: freeze the locations and widths of both pillars' components and
move only their weights, along a straight line in weight space. The path stays in a
fixed family of at most $2N$ components, is arbitrage-free whenever the two pillars
are, and for full-support kernels is realized by a unique continuous local-volatility
diffusion. The same construction chains across a term structure at additive cost, 
which is what makes it practical: Section~\ref{sec:applications} uses it to lift the 
Brigo--Mercurio mixture model to time-varying weights, and to avoid the tensor 
blow-up that arises in one proposed extension of SANOS.

\section{Related work}

Two mixture-dynamics models sit closest to this construction: Brigo--Mercurio
\cite{brigomercurio2002} and SANOS \cite{sanos2026}, which we engage in detail in
Section~\ref{sec:applications}. The Bass local-volatility model
\cite{conzehl2021,bass2020} produces the canonical smoothest arbitrage-free
interpolant but leaves the mixture family. The present time-varying-weight, mixture-preserving
case is covered cleanly by none of these; the validity theory rests on Kellerer
\cite{kellerer1972}, the peacock framework \cite{hpry2011}, and Lowther's
diffusion characterization \cite{lowther2008,lowther2009}, with Dupire
\cite{dupire1994} supplying the local-volatility formula.

\section{Setup}

Let the forward $X_t$ be a martingale ($\E[X_t]=F$ for all $t$). Its risk-neutral
density is given only at a finite set of \emph{pillar} maturities
$t_0<t_1<\dots<t_n$. At each pillar $t_k$ ($k=0,\dots,n$) the density is an
$N$-component mixture,
\begin{equation}
  p_{t_k}(x) \;=\; \sum_{i=1}^{N} w_i(t_k)\,\phi\!\big(x;\mu_i(t_k),\sigma_i(t_k)\big),
  \qquad \sum_i w_i(t_k)=1,\quad \sum_i w_i(t_k)\mu_i(t_k)=F,
  \label{eq:mix}
\end{equation}
with $\phi$ a fixed kernel (here Gaussian; the lognormal case is analogous with
Black--Scholes terms) and $N$ the number of components. We write $\mathcal{M}_k$ for
the family of mean-$F$ mixtures of at most $k$ such kernels, so each
$p_{t_k}\in\mathcal{M}_N$. The subscript $t_k$ is a
\emph{discrete} label: these $n+1$ snapshots are all we observe. The task is to fill
in the density at every intermediate time, that is, to build a \emph{continuous-time}
family $(p_t)_{t\in[t_0,t_n]}$ that matches the data at the pillars,
$p_t\big|_{t=t_k}=p_{t_k}$, by interpolating the mixture parameters
$(w_i(t),\mu_i(t),\sigma_i(t))$ between consecutive pillars.
Section~\ref{sec:construction} constructs such a family, continuous in $t$ and inside
the mixture family.

\begin{definition}[Peacock]
The time-indexed collection $(p_t)_{t\in[t_0,t_n]}$, all members sharing the mean
$F$, is a \emph{peacock} if it spreads out over time:
for any two dates $s\le t$ and every convex function $\varphi$,
\begin{equation*}
  \E\big[\varphi(X_s)\big] \;\le\; \E\big[\varphi(X_t)\big],
  \qquad
  \E\big[\varphi(X_t)\big]=\int\varphi(x)\,p_t(x)\,dx .
\end{equation*}
Equivalently, $p_t$ is a \emph{mean-preserving spread} of $p_s$: same mean, more
dispersion. This ordering is the \emph{convex order}, written $p_s\cx p_t$.
\end{definition}

Writing $C(t,K)=\E[(X_t-K)^+]$ for the call function and $p_t=\partial_{KK}C$ for
the density, the peacock property is equivalent to the absence of calendar-spread
arbitrage,
\begin{equation}
  \partial_t C(t,K)\;\ge\;0\qquad\text{for all }K.
\end{equation}

\section{Validity: peacock, Fokker--Planck, and local volatility}

A continuous martingale matching the marginals solves
$dX_t=\sigma_{\mathrm{loc}}(t,X_t)\,dW_t$, and the Fokker--Planck (forward Kolmogorov)
equation $\partial_t p_t=\tfrac12\partial_{xx}(a\,p_t)$ inverts to Dupire's formula
\begin{equation}
  a(t,K)\;:=\;\sigma_{\mathrm{loc}}^2(t,K)\;=\;\frac{2\,\partial_t C(t,K)}{p_t(K)}.
  \label{eq:dupire}
\end{equation}
Since $p_t>0$, the sign of $a$ equals the sign of $\partial_tC$, so
\begin{equation}
  \text{peacock}\;\Longleftrightarrow\;\partial_tC\ge 0\;\Longleftrightarrow\;a\ge 0 .
\end{equation}
The existence of an actual diffusion is more delicate. Kellerer's theorem
\cite{kellerer1972} (see also \cite{strassen1965,hpry2011}) guarantees \emph{a}
Markov martingale with the given marginals; whether it can be taken to be a
diffusion is a separate question. The result we need is due to Lowther
\cite{lowther2008,lowther2009}; we quote the form stated in
\cite{bachelier2dupire2021}.

\begin{theorem}[Lowther]
\label{thm:lowther}
Let $(p_t)_{t\ge0}$ be a peacock such that $t\mapsto p_t$ is weakly continuous
and each $p_t$ has convex support. Then there is a \emph{unique} strongly
continuous Markov martingale $X$ with $X_t\sim p_t$.
\end{theorem}

A continuous strong-Markov martingale in one dimension is, in the regular case, a
local-volatility diffusion, so Theorem~\ref{thm:lowther} makes ``peacock $=$ local volatility'' precise
under two hypotheses: weak time-continuity and \emph{convex
(connected) support}. The second is the one that matters here.

\begin{proposition}[Full-support kernels]
\label{prop:fullsupport}
If $\phi$ has full support (Gaussian on $\R$, lognormal on $\R_+$), then every
mixture $p_t$ has support equal to $\R$ (resp.\ $\R_+$), which is convex. Hence
for a weakly continuous peacock of such mixtures, Theorem~\ref{thm:lowther} applies
and a unique continuous local-volatility diffusion exists, including in the
multimodal case.
\end{proposition}

The only way to violate the convex-support hypothesis is a genuinely
\emph{disconnected} support (a true gap of zero density that probability must cross);
this is where a pure diffusion fails and jumps become necessary. Full-support kernels
never produce this, so in our setting validity reduces exactly to the peacock check.

\section{A $2N$-component construction}
\label{sec:construction}

Take the \emph{frozen pool}: the union of both endpoints'
components, with their locations and widths held fixed for all $t$,
$\{\phi_j\}_{j=1}^{2N}$. Only the weights $w(t)\in\R^{2N}$ move; we call a pool
component \emph{active} at time $t$ when it carries positive weight, $w_j(t)>0$. The
call function is
linear in the weights, $C(t,K)=\sum_j w_j(t)\,g_j(K)$ with $g_j$ the single-component
call, so $\partial_t C=\sum_j \dot w_j\,g_j$. The weight changes $v=\dot w$ that keep
the path arbitrage-free form a convex cone,
\begin{equation}
  \mathcal{C}=\Big\{v:\textstyle\sum_j v_j=0,\;\sum_j v_j\mu_j=0,\;\sum_j v_j g_j(K)\ge0\;\forall K\Big\},
\end{equation}
i.e.\ keep total weight at $1$, keep the mean at $F$, and add no calendar arbitrage.
The two endpoints sit on opposite faces of the simplex; after pinning the mean, the
space of admissible weight configurations has dimension $2N-2$.

The two pillars need not have the same number of components. If $m_0$ has $N_0$ and
$m_1$ has $N_1$ components, the frozen pool holds $N_0+N_1$ of them, and the counts
above read $N_0+N_1$ and $N_0+N_1-2$ in place of $2N$ and $2N-2$. Only the
\emph{weights} are interpolated, so the number of \emph{active} components is free to
change along the path: at $t_0$ exactly $m_0$'s components are active (the rest carry
zero weight), at $t_1$ exactly $m_1$'s are, and in between the chord generically
activates the whole pool. A single component splitting into two ($N_0=1,N_1=2$) thus
has active count $1\to(3)\to2$, and a bimodal pair merging into one ($N_0=2,N_1=1$)
runs $2\to(3)\to1$: the interior is a three-component mixture even though both
endpoints are smaller, and the endpoints are recovered exactly because the absent
components sit at zero weight. The equal-count case $N_0=N_1=N$ is the $2N$ pool named
in the section title.

\begin{proposition}[The chord always works]
\label{prop:backstop}
For convex-ordered endpoints, the straight line in weight space
$w(t)=(1-s(t))\,w^{(0)}+s(t)\,w^{(1)}$ with $s$ increasing, $s(t_0)=0,\,s(t_1)=1$,
gives $C(t,K)=(1-s)C_0(K)+s\,C_1(K)$. Then:
(a) the mean is preserved for all $t$; (b) $\partial_tC=\dot s\,(C_1-C_0)\ge0$, so the
path is a peacock; (c) $p_t=(1-s)p_0+s\,p_1>0$, so $a$ in \eqref{eq:dupire} is finite.
The path is continuous in $t$ (the weights move continuously with $s$), stays in the
$2N$ family, and subdividing further on the same pool keeps the count at $2N$, never
growing to $3N,4N,\dots$.
\end{proposition}

So mixture-preserving arbitrage-free interpolation always exists in $\mathcal{M}_{2N}$, with no
conditions beyond the endpoints themselves being arbitrage-free. The doubling from $N$
to $2N$ is paid once. The cone $\mathcal{C}$ of admissible velocities is itself
$(2N-2)$-dimensional, and the chord is one ray inside it, so beyond the chord there are
$2N-3$ transverse directions; they are one-directional, since $\mathcal{C}$ is a cone
and reversing any of them reintroduces calendar arbitrage. These bow the path without
moving the endpoints, and can be used to line up the local-volatility slice on the two
sides of a pillar. Figure~\ref{fig:algo} collects the construction as a recipe.

\begin{figure}[t]
\centering
\fbox{\begin{minipage}{0.93\textwidth}
\textbf{Algorithm.} In-family arbitrage-free interpolation between two pillars.\\[2pt]
\textbf{Input:} mixtures $m_0,m_1$ with the same forward mean $F$; pillar times
$t_0<t_1$; a strike grid.\\[3pt]
\begin{enumerate}\setlength{\itemsep}{1pt}
\item \textbf{Pool.} Concatenate the components: locations $\mu_j$ and widths
$\sigma_j$ are
the union of $m_0$'s and $m_1$'s, frozen. Set
$w^{(0)}=(\,m_0\text{'s weights},\,0\dots)$ and $w^{(1)}=(0\dots,\,m_1\text{'s
weights})$ on this pool.
\item \textbf{No-arb check.} On the strike grid require $C_1(K)\ge C_0(K)$ for all $K$
(peacock / no calendar-spread arbitrage). If it fails, the two pillars are themselves
inconsistent; stop.
\item \textbf{Clock.} Pick an increasing $s(t)$ with $s(t_0)=0,\,s(t_1)=1$ (the
identity, or a \emph{variance clock} $s(t)=\tfrac{\Sigma(t)-\Sigma_0}{\Sigma_1-\Sigma_0}$,
where $\Sigma(t)=\Var(X_t)$ is the mixture variance and $\Sigma_k=\Sigma(t_k)$).
\item \textbf{Weights.} $w(t)=(1-s)\,w^{(0)}+s\,w^{(1)}$. (Optionally add a lateral
move from the null space of step 1's two constraints to match a pillar slice.)
\item \textbf{Density.} $p_t(x)=\sum_j w_j(t)\,\phi(x;\mu_j,\sigma_j)$.
\item \textbf{Local vol.} $a(t,K)=\dfrac{2\,\partial_t C}{p_t(K)}=\dfrac{2\,\dot
s\,(C_1(K)-C_0(K))}{p_t(K)}$, \ $\sigma_{\mathrm{loc}}=\sqrt a$.
\end{enumerate}
\textbf{Output:} a continuous family $\{p_t\}_{t\in[t_0,t_1]}$ in the $2N$ family with
$a\ge0$ everywhere. Chain over $t_0,t_1,\dots,t_n$ to cover all pillars.
\end{minipage}}
\caption{The construction of Proposition~\ref{prop:backstop} as a recipe.}
\label{fig:algo}
\end{figure}

\subsection{Examples}

\begin{figure}[t]
\centering
\includegraphics[width=\textwidth]{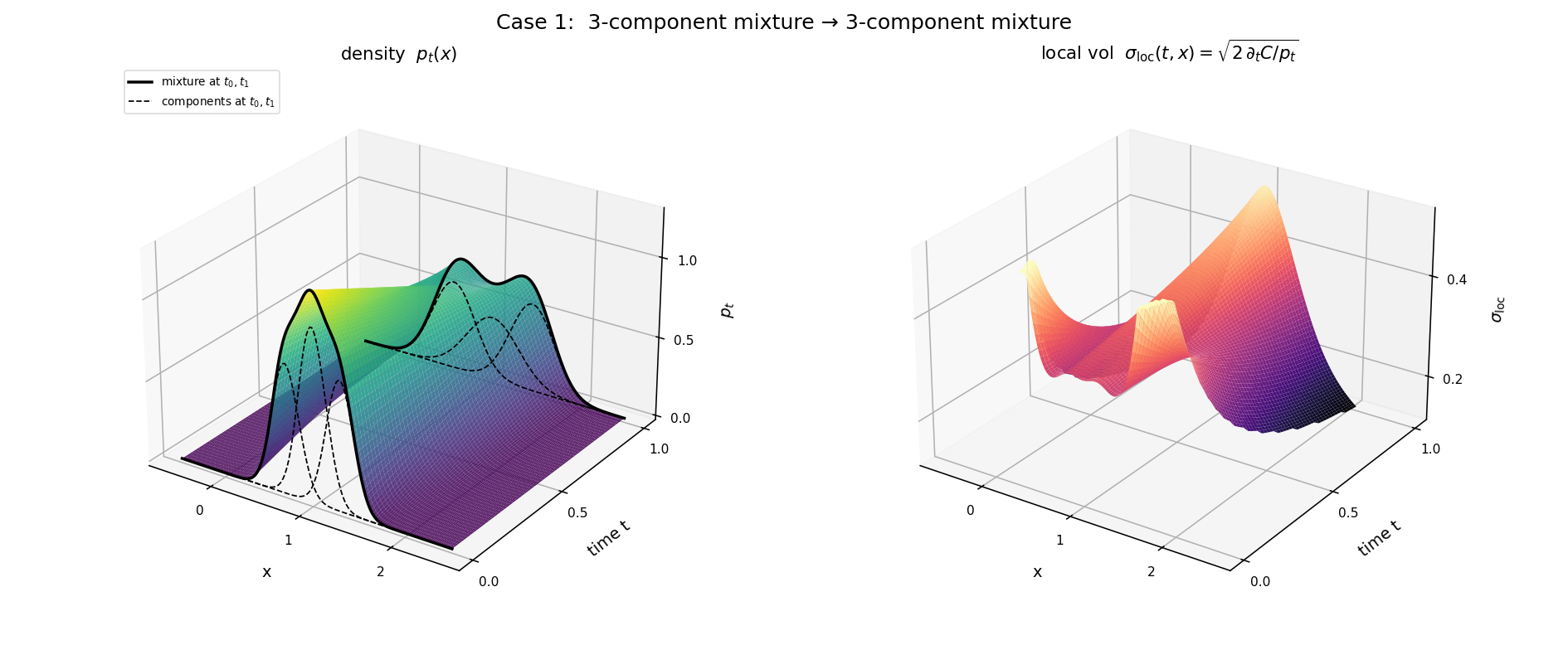}
\caption{\textbf{Case 1.} A three-component mixture relaxing into another
three-component mixture with shifted means and wider components.}
\label{fig:case1}
\end{figure}

\begin{figure}[t]
\centering
\includegraphics[width=\textwidth]{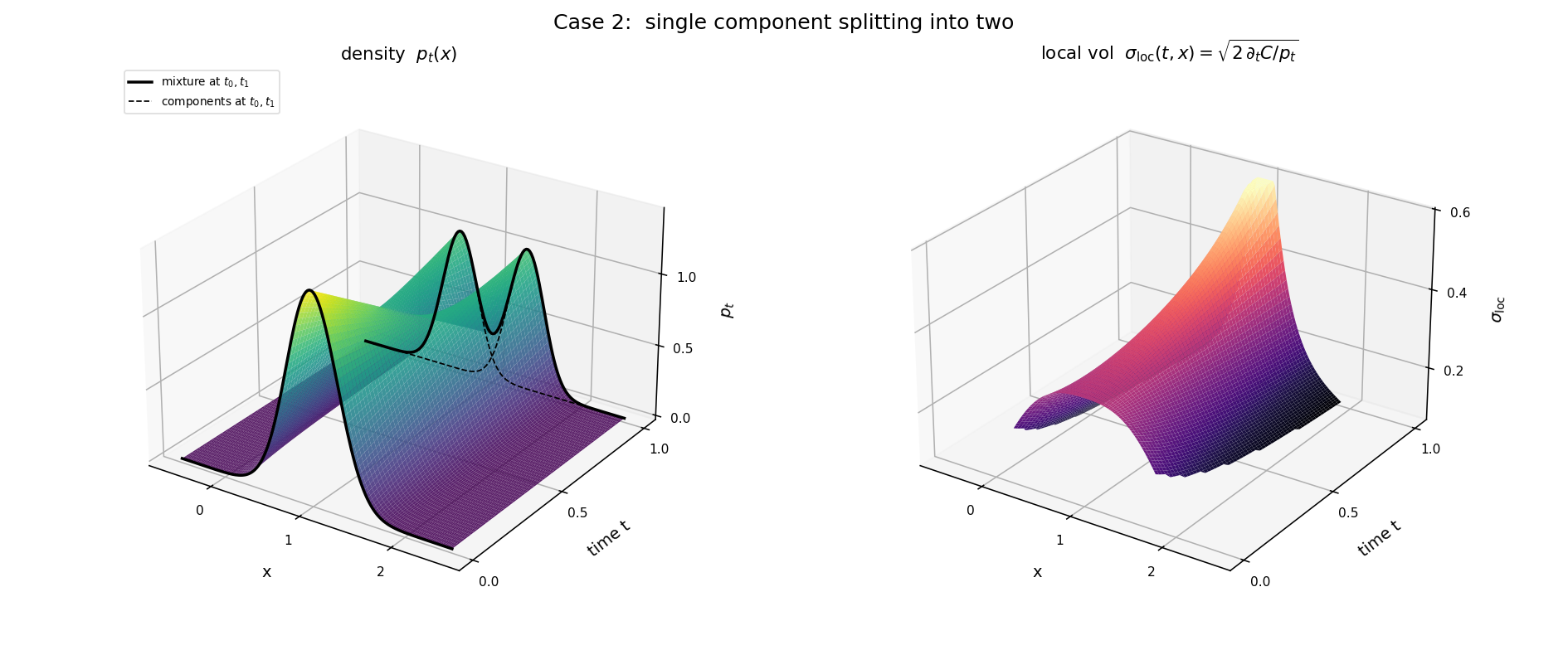}
\caption{\textbf{Case 2.} A single component splitting into two, the mean fixed.
Active-component count $1\to(3)\to2$: the frozen pool holds three components, two of
them carrying zero weight at each endpoint.}
\label{fig:case2}
\end{figure}

\begin{figure}[t]
\centering
\includegraphics[width=\textwidth]{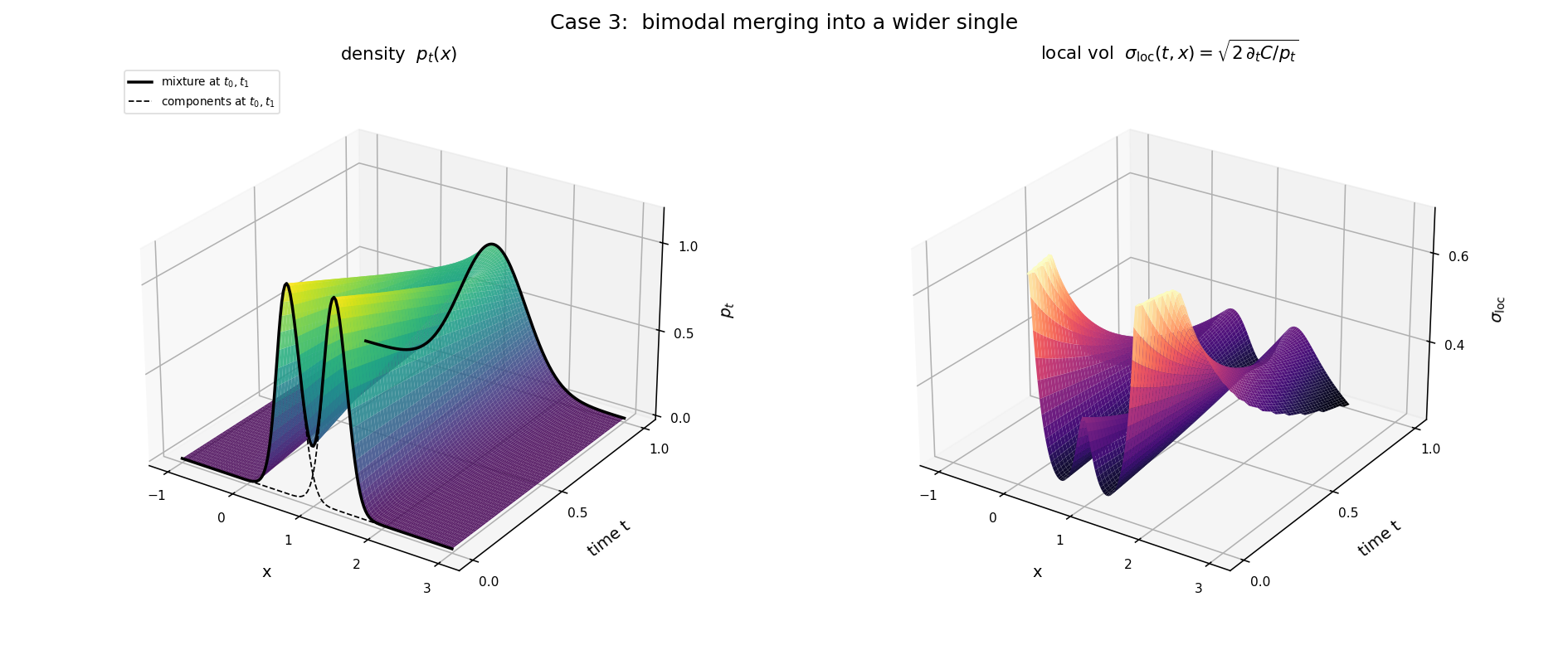}
\caption{\textbf{Case 3.} A bimodal density diffusing into a wider single mode.
Active-component count $2\to(3)\to1$. The local-vol ridge over the valley is the
conditioning effect of Section~\ref{sec:cond}.}
\label{fig:case3}
\end{figure}

Figures~\ref{fig:case1}--\ref{fig:case3} run the algorithm on three representative
moves between two pillars (mean fixed at $F=1$). Each figure shows the interpolated
density $p_t(x)$ as it evolves (left) next to the implied local-volatility surface
$\sigma_{\mathrm{loc}}(t,x)$ (right), in the same orientation. All three pass the
peacock check, so each is a valid local-volatility diffusion. Case~3 shows the
conditioning issue of Section~\ref{sec:cond}: the local volatility spikes along the
low-density valley while the two modes are still separated.

\clearpage

\subsection{Conditioning in deep bimodal regimes}
\label{sec:cond}

By Proposition~\ref{prop:fullsupport} the local volatility always exists for our
kernels, but it can still be badly conditioned. When the endpoints are strongly
bimodal with a thin valley between the modes, the numerator $\partial_t C$ stays
bounded away from zero in the valley (heuristically, because probability mass must
cross it), while the denominator $p_t$ there decays exponentially in the
mode separation. By \eqref{eq:dupire}, $a=2\partial_tC/p_t$ is then finite but very
large: the local-volatility slice develops a sharp spike over the valley. As the
separation grows the support stays connected (so Theorem~\ref{thm:lowther} still
applies) and $a$ diverges only in the limit of truly disjoint support, the boundary at
which jumps would become mandatory. Case~3 (Figure~\ref{fig:case3}) shows this as the
tall ridge in the local-volatility surface over the valley.

\section{Applications to mixture term-structure models}
\label{sec:applications}

Several option-surface models build each expiry from a (log)normal mixture and impose
convex order across expiries, exactly the setting of this paper. For them the
frozen-pool construction is directly useful. Everything transfers to the lognormal
case: take $\phi$ the Black--Scholes density, $g_j$ the Black--Scholes call, and the
mean constraint $\sum_i w_i\mu_i=F$ the forward; by
Proposition~\ref{prop:fullsupport} a lognormal mixture has support $\R_+$, which is
convex, so validity again reduces to the peacock check. We give two applications.

\subsection{Time-varying weights, and a strong-solution gap, in Brigo--Mercurio}

The lognormal-mixture local-volatility model of Brigo--Mercurio
\cite{brigomercurio2002} keeps the mixture \emph{weights} static and grows the
component variances. Calibrating each expiry independently and joining the
calibrations with the frozen-pool chord lifts this to \emph{time-varying} weights
while staying lognormal-mixture and arbitrage-free.

Their Section~4.1 (``shifting each basic distribution'') adds per-component shifts
$\beta_i$, giving a local-volatility SDE for which they write that ``it is still to be
verified that\dots the corresponding SDE\dots has a unique strong solution.'' The
shifted components have supports $[\beta_i e^{\mu t},\infty)$; their union is a
connected half-line, so the convex-support hypothesis of Theorem~\ref{thm:lowther}
holds. Hence, as soon as the calibrated marginals are convex-ordered in $T$ (a
cheap, checkable condition), Lowther's theorem supplies a unique continuous
strong-Markov martingale with those marginals, i.e.\ the mimicking local-volatility
diffusion exists and is unique in law.\footnote{Lowther gives uniqueness in law of a
continuous strong-Markov martingale matching the marginals; this settles existence
and uniqueness of the mimicking diffusion, marginally weaker than pathwise strong
uniqueness for one fixed SDE.} For instance, a two-expiry shifted-lognormal example
with weights moving $0.50/0.50\to0.35/0.65$ has connected support, a non-negative
peacock margin, and a non-negative local variance throughout.

\subsection{Free per-strike widths at additive cost}
In SANOS \cite{sanos2026} each expiry's call surface is a Black--Scholes-call
mixture, $\hat C(T_j,K)=\sum_{i=1}^{N_j} q_j^i\,\mathrm{Call}(K_j^i,K,V_j)$, with the
weights $q_j$ read as a discrete transition density, an $N_j$-component mixture in
our sense. The base model is thus already in our setting. To free the per-strike
\emph{width} as well, the authors propose (their Theorem~3.9) representing the surface
as the law of a \emph{product} martingale $Z_j=\prod_{\ell\le j}X_\ell Y_\ell^{X_\ell}$.
This buys the extra generality at a price the authors themselves flag as prohibitive:
pricing the $T_j$ marginal needs the full tensor structure (their Prop.~3.10), a sum
over multi-indices $(i_1,\dots,i_j)$ and so $\prod_{k\le j}N_k$ components, with the
flow no longer Markov (their Remark~3.12).

The frozen pool reaches the same generality with a sum instead of a product. Chain the
chord across the expiries $T_0,\dots,T_M$: on each interval $[T_l,T_{l+1}]$ it
interpolates the two pillar mixtures from their $N_l+N_{l+1}$ pooled components, so the
whole chain uses $\sum_l N_l$ components in total and at most $N_l+N_{l+1}$ at any one
time. The cost is additive in the number of expiries, against the product's
$\prod_l N_l$. And the result is still a genuine diffusion: convex order is transitive,
so the concatenated flow is a peacock, its support is connected for full-support
kernels, and Lowther's theorem (Theorem~\ref{thm:lowther}) realizes it as a unique
continuous Markov local-volatility diffusion. The product martingale, by contrast, is
not Markov.

For $N_l\equiv N$ this is $N^{M}$ versus $NM$; Table~\ref{tab:tensor} shows the gap for
$N=5$. There is a trade-off the other way: the SANOS product is automatically a
peacock, so it needs no convex-order check, while the frozen pool does. But that check
is a \emph{one-time} test on the pillar data, the linear comparison $C_{l+1}\ge C_l$ on
each consecutive pair of marginals, which any arbitrage-free model has to pass anyway.
Once a pair passes, Proposition~\ref{prop:backstop} makes the chord between them
arbitrage-free at every intermediate $t$, so nothing more is checked during the
interpolation.

\begin{table}[H]
\centering
\begin{tabular}{r|r|l}
expiry $T_l$ & SANOS $\prod_{k\le l}N_k$ & frozen pool ($N=5$ per expiry)\\\hline
1 & \phantom{000}5 & 10 per interval, \phantom{0}5 total\\
2 & \phantom{00}25 & 10 per interval, 10 total\\
3 & \phantom{0}125 & 10 per interval, 15 total\\
4 & \phantom{0}625 & 10 per interval, 20 total\\
5 & 3125           & 10 per interval, 25 total\\
\end{tabular}
\caption{Component count for a chain of $M=5$ expiries with $N=5$
free-per-strike-width components each: the SANOS tensor $\prod_k N_k$ against the
frozen-pool sum $\sum_k N_k$ (at most $N_l+N_{l+1}$ active at once).}
\label{tab:tensor}
\end{table}

\section{Conclusion and further work}

Mixture-preserving, arbitrage-free interpolation always exists in a fixed
$2N$-component family, with no condition beyond the two pillars being arbitrage-free,
and for full-support kernels it is a unique continuous local-volatility diffusion.
Chained across a term structure its cost is additive, $\sum_l N_l$ components in all.

The chord of Proposition~\ref{prop:backstop} is the simplest such interpolation, not
the only one. It moves the pooled weights along a straight line with the component
locations and widths frozen, but neither is forced: the weights can follow any path in
the admissible cone $\mathcal{C}$, whose $2N-3$ transverse directions
Section~\ref{sec:construction} noted, and the locations and widths can drift in time as
well, as long as the marginals stay convex-ordered.

This freedom leaves room to choose the interpolation by a criterion rather than take
the chord by default, for example the smoothest local-volatility surface among the
mixture-preserving peacocks through the data. That is the mixture-preserving
counterpart of the Bass interpolant \cite{conzehl2021,bass2020}, which is the smoothest
arbitrage-free interpolant but leaves the mixture family. Whether the count can be
dropped from $2N$ to $N$ is also open: in every example we tried an $N$-component
mixture-preserving peacock path could be found, typically by inflating component
variances while rearranging weights and locations, though we have neither a proof nor a
counterexample, the obstruction being that the $N$-component family is non-convex (the
chord leaves $\mathcal{M}_N$ for $\mathcal{M}_{2N}$).

None of this crosses the support boundary of Section~\ref{sec:cond}. A diffusion stays
valid up to the point where the modes separate into disjoint support; past it no
local-volatility surface exists. The marginals there are still a peacock, so by
Kellerer's theorem \cite{kellerer1972} a Markov martingale matching them still exists;
it simply can no longer be continuous, and must jump across the gap. The natural
extension is to that wider class, a Markov jump-diffusion, coupling the frozen pool to
a jump component beyond the disconnection.

\end{document}